\documentclass[twocolumn]{aastex631}

\shorttitle{Hierarchical absorption in LAMOST normalized spectra}
\shortauthors{Shen et al.}
\graphicspath{{./}{figures/}}

\begin{document}

\title{Hierarchical Absorption in LAMOST low Resolution Normalized Spectra}

\author[0000-0003-4445-6504]{Yu-Fu Shen}
\affiliation{Changchun Observatory, National Astronomical Observatories, Chinese Academy of Sciences, Jingyuetan National Scenic Area, Changchun, 130117, China}

\author[0000-0002-1126-9289]{Zhuohan Li}
\affiliation{CAS Key Laboratory of Optical Astronomy, National Astronomical Observatories, Chinese Academy of Sciences, Beijing, 100101, China}
\affiliation{School of Astronomy and Space Science, University of Chinese Academy of Sciences, Beijing, 100049, China}



\begin{abstract}

According to the hierarchical clustering scenario, galaxies like the Milky Way form hierarchically, and many supporting evidences have been found in the Galactic halo. However, most stars in the Milky Way are disk stars. Disk stars have almost lost their spatial distribution and kinematic features at birth, retaining solely chemical signatures. Identifying such substructures using abundances of iron or light elements is difficult due to their high degeneracy. Heavy elements, especially neutron capture elements have limited sources so have lower degeneracy, but spectral line fitting of these elements is tough, requiring mid to high resolution spectra, which are currently limited in sample size. This work utilizes the collective effect of many spectral lines from several elements, especially neutron capture elements to weaken the degeneracy of [Fe/H]. The analysis suggests the presence of at least 12 clusters, due to the hierarchical absorption in LAMOST low resolution normalized Spectra. More detailed work is needed to ascertain whether this hierarchical absorption is natural.

\end{abstract}



\section{Introduction} \label{sec:intro}

According to the hierarchical clustering scenario, galaxies are assembled by merging and accretion of many clusters of different sizes and masses\citep{Klypin_1999}. Cold dark matter model also predicts that the galaxies like the Milky Way form hierarchically\citep{2005ApJ...635..931B,Komatsu_2011,annurev:/content/journals/10.1146/annurev-astro-091916-055313}. However, the disk stars, which account for the vast majority of the Galaxy, appear to be continuous. The interaction between stars and the Galaxy could be the explanation of the contradiction. A star cluster in or near the Galaxy initially had clustering properties; under the interaction with the Galaxy, it gradually disintegrates and forms a stellar stream\citep{2002MNRAS.332..915I,2002ApJ...570..656J,2012ApJ...748...20C,2021ApJ...914..123I}; then it will lose spatial distribution characteristics and only retain kinematic features, such as Gaia Sausage\cite{2018MNRAS.478..611B}; finally, the kinematic features of the stars will be lost, leaving only chemical information. The above process has been verified in the galactic halo. As for the disk, some works suggest that a significant fraction of the Galactic Disk also formed hierarchically\citep{2002ApJ...574L..39G,2003ApJ...591..499A}, but the evidence is limited. If the Galactic disk is indeed formed hierarchically, it is believed that most of the clusters in the disk are only recognizable by their element abundances.

The elemental abundances obtained by spectral line fitting are difficult to use for distinguishing potential substructures within the Galactic disk. On the one hand, the elemental abundances derived in this way inherently contain errors, leading to a broadening of the abundance distribution. For low-resolution spectra, the unclear spectral line profiles introduce errors. For medium-to-high-resolution spectra, errors can be introduced by factors such as the difficulty in determining the continuum spectrum, uncertainties in atmosphere parameters and atomic parameters, and non-local thermodynamic equilibrium (NLTE) effects in metal-poor stars\citep[e.g.][]{2016MNRAS.463.1518A,2022ApJ...931..147L,2022A&A...665A..10L,Shen_2023}. However, even if there were ways to significantly improve the accuracy of abundances derived from spectral line fitting, it may still not be possible to identify clusters within the Galactic disk if we only focus on Fe and light elements, due to the high degeneracy of these elements\citep[e.g.][]{2023MNRAS.520.5671H}. Abundant sources of elements lead to higher degeneracy. Almost all the stars can generate iron and light elements. In contrast, the sources of heavy elements, especially neutron capture elements, are much fewer. The neutron capture process contains rapid neutron capture process (r-process) and slow neutron capture process (s-process). For elements heavier than Fe (or Zn), based on current understanding, they can only produced by neutron capture process. The astrophysical sites of the r-process have been debated, such as type II supernovae, binary neutron star mergers, cataclysmic variable stars, and other white dwarf systems\citep{1974ApJ...192L.145L,1999A&A...341..499R,2007A&A...467.1227A,2008ApJ...676L.127H,2010A&A...517A..80F,2011ApJ...726L..15W,2011ApJ...731....5A,Goriely_2011,Wanajo_2013,2013ApJ...770L..22W,2014ApJ...789L..39W}. Among them, only the binary neutron star merger is confirmed till now\citep{Chornock_2017,2017Sci...358.1570D,doi:10.1126/science.aaq0073,2017Natur.551...67P,2017PhRvL.119p1101A,2017Natur.551...75S,2017ApJ...848L..24V,2017Sci...358.1574S,2017ApJ...848L..13A,2019Natur.574..497W}. The s-process occurs in the late evolution stage of small and medium-mass stars, but due to the uncertainty of the reaction rate of neutron source reactions, theoretical models cannot well constrain element yields\citep{1999ARA&A..37..239B,2005ARA&A..43..435H,2014PASA...31...30K,2010ApJ...710.1557P,10.1093/mnras/stv2723,refId0,Limongi_2018}. The abundance of the neutron capture process elements in stars is necessary for in-depth understanding of the r-process and s-process, as well as the analysis of the substructures within the Galactic disk.

The measurement of the abundance of neutron capture process elements is difficult because fitting their spectral lines requires mid or high spectral resolution. So, the sample size of stars with s-process abundances\citep{1988A&A...198..187J,1990ApJ...352..709M,1999A&A...345..127V,2018A&A...620A.148S,2020A&A...635L...6S} and r-process abundances\citep{2018ARNPS..68..237F,2018ApJ...858...92H,2018ApJ...865..129R,2018ApJ...868..110S,2018ApJ...854L..20S,2019ApJ...874..148S,Ezzeddine_2020,2020ApJS..249...30H} is small. Accurate abundance measurements must fit the spectral lines, but rough abundance estimate sometimes works in photometric data or low-resolution spectra (R $\sim$2000). There have been successful methods to estimate [Fe/H] using photometry\citep[e.g.][]{2023ApJ...957...65H} or low-resolution spectra\citep[e.g.][]{2011RAA....11..924W}. As for neutron capture process elements, if the abundance of them is obviously higher than average (so called s-type stars or r-type stars), photometric data and low-resolution spectra can help to identify them\citep{1993A&A...271..463J,1998A&A...333..613C,2002A&A387,2006AJ....132.1468Y,Chen_2019,Chen_2023}, but hard to provide the abundance. Besides, the low-resolution spectra still have potential. Even though there are only a few spectral lines that can be fitted in low-resolution spectra, it still retains a lot of information about elements\citep{2017ApJ...843...32T}. If the overall effect of all spectral lines of a certain element can be identified on low-resolution spectra, abundance measurement can be achieved. However, even with this method, only few heavy elements can be measured\citep{2019ApJS..245...34X}, because the abundances of many heavy elements are too low, or the distribution of the absorption lines of several heavy elements are too close.

It is difficult to estimate the abundance of a single heavy element, but it should be easier to estimate the overall abundance of several heavy elements. The s- and r-processes always form a series of elements at the same time, so it is reasonable to regard all neutron capture elements as a whole. Most of the absorption lines of s-process elements and r-process elements are more concentrated in the violet end of the optical spectrum, so the effects of them in low-resolution spectra are close, which will be shown in this paper.

This paper is organized as follows. Section \ref{data} introduces the synthetic and observed data used in this work. Section \ref{met} provides the detailed process for obtaining evidence of the Hierarchical absorption in LAMOST low resolution normalized spectra. Section \ref{dis} mainly provides a simulation, and some other discussions.

\section{Data}\label{data}

\subsection{Synthetic spectra}

\begin{figure}
\centering
\includegraphics[width=0.9\linewidth]{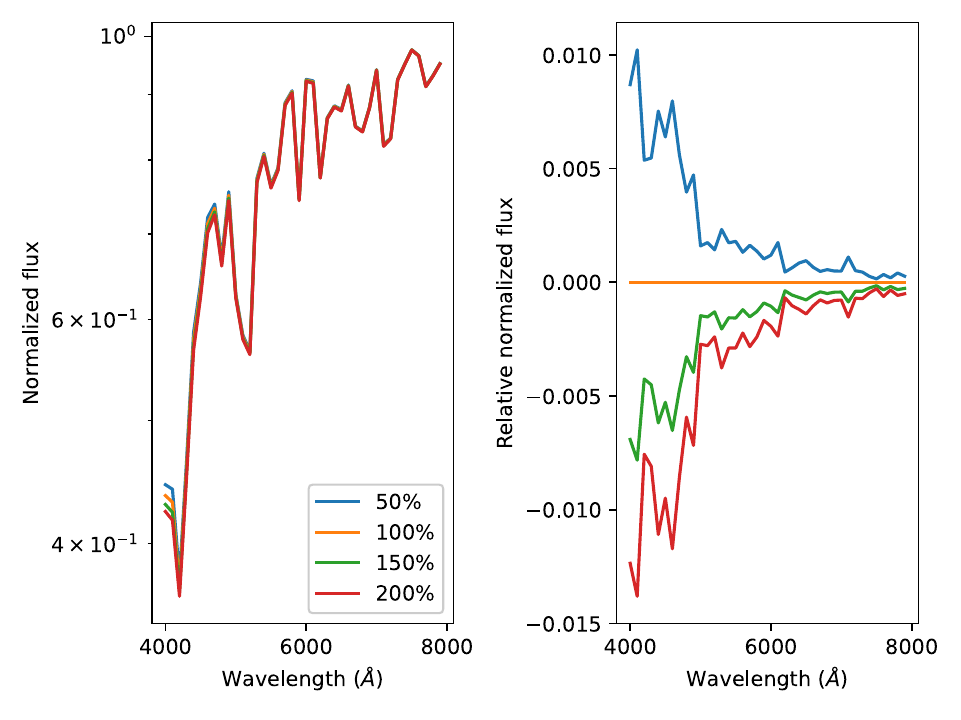}
\caption{The left panel is the synthetic normalized spectra with atmospheric parameters $T_\mathrm{eff}=4000 \text{K}$, $\log g = 4.0$, $\mathrm{[Fe/H]}=0.0$. The percentage in the legend means the proportion of neutron capture elements of solar abundance. The right panel is the left panel subtracted by the 100\% line. This graph qualitatively demonstrates the sensitivity of various wavelengths in the spectrum to the neutron capture element abundance.}\label{1}
\end{figure}

\begin{figure*}
    \centering
    \includegraphics[width=0.9\linewidth]{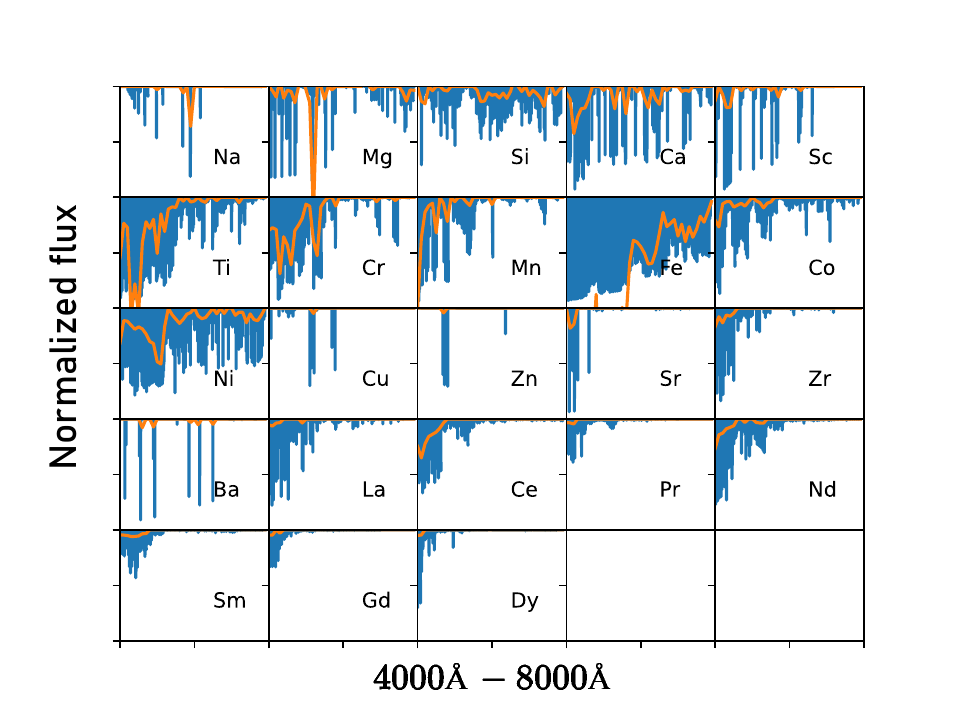}
    \caption{The distribution of spectral lines in synthetic solar spectra. The blue spectra R$\sim$300,000, the orange spectra R$\sim$60. The y axis of all panels are set between 0 and 1. The absorption depth of the orange spectra are enlarged 30 times.}
    \label{manyelements}
\end{figure*}

Using MARCS atmosphere models\citep{2008A&A...486..951G} and SynthV code\citep{1996ASPC..108..198T,2019ASPC..518..247T}, synthetic stellar spectra can be generated if abundances of elements are given. Fig.~\ref{1} shows the change in the spectrum after modifying only the abundances of $Z>30$ elements with the atmospheric model unchanged. Fig.~\ref{manyelements} are the spectra only contain one element, they are used to show the distribution of absorption lines of elements. The elements that do not have obvious absorption lines between $4000 \text{\AA}$ and $8000 \text{\AA}$ in the synthetic solar spectra are not included in Fig.~\ref{manyelements}.

\subsection{Observed data}

This work analyses the low-resolution spectra (R$\sim$1800) from the Large Sky Area Multi-Object Fiber Spectroscopic Telescope \citep[LAMOST][]{2012RAA....12.1197C,Zhao_2012,Liu_2015,YAN2022100224}. 5,096,468 deduplicated stars from the LAMOST LRS Stellar Parameter Catalog of A, F, G and K Stars\citep{2012RAA....12..723Z,2011RAA....11..924W} are considered. After removing the giants and several other processes, 3,614,633 stars are analyzed, the details are introduced in the next section. Based on the proper motion, parallax, and radial velocity provided by Gaia DR3\citep{2016A&A...595A...1G,2023A&A...674A...1G} and LAMOST, Galactic space-velocity components\citep{1987AJ.....93..864J}, their form in Galactocentric Cartesian coordinate system\citep{Wu_2022}, and the orbital elements under McMillan17 gravitational potential\citep{2017MNRAS.465...76M} are calculated to make sure the clusters found do not have obvious kinematic features.

\section{Method}\label{met}

\begin{figure}
    \centering
    \includegraphics[width=0.9\linewidth]{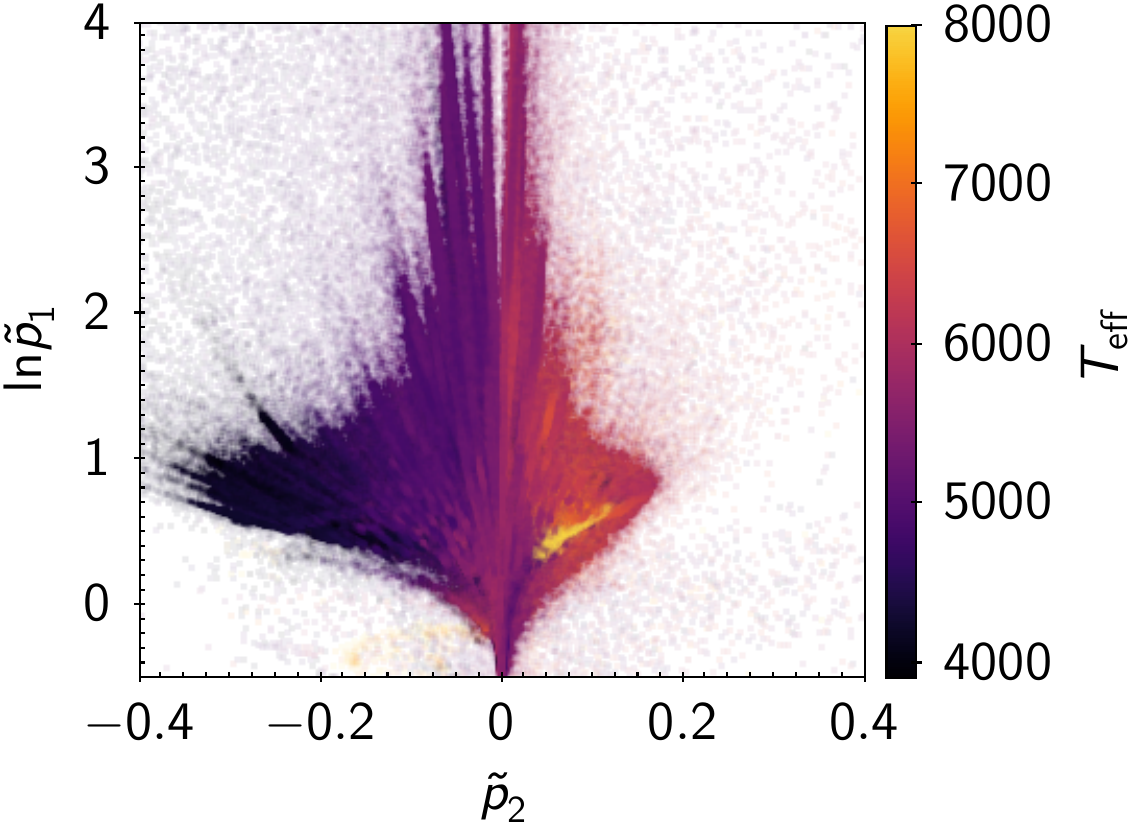}
    \caption{The figure primarily depicts the central distribution of the fitting parameters for the entire deduplicated LAMOST LRS Stellar Parameter Catalog of A, F, G and K Stars. The colors represent $T_\mathrm{eff}$.}
    \label{allteff}
\end{figure}

\begin{figure*}
    \centering
    \includegraphics[width=0.45\linewidth]{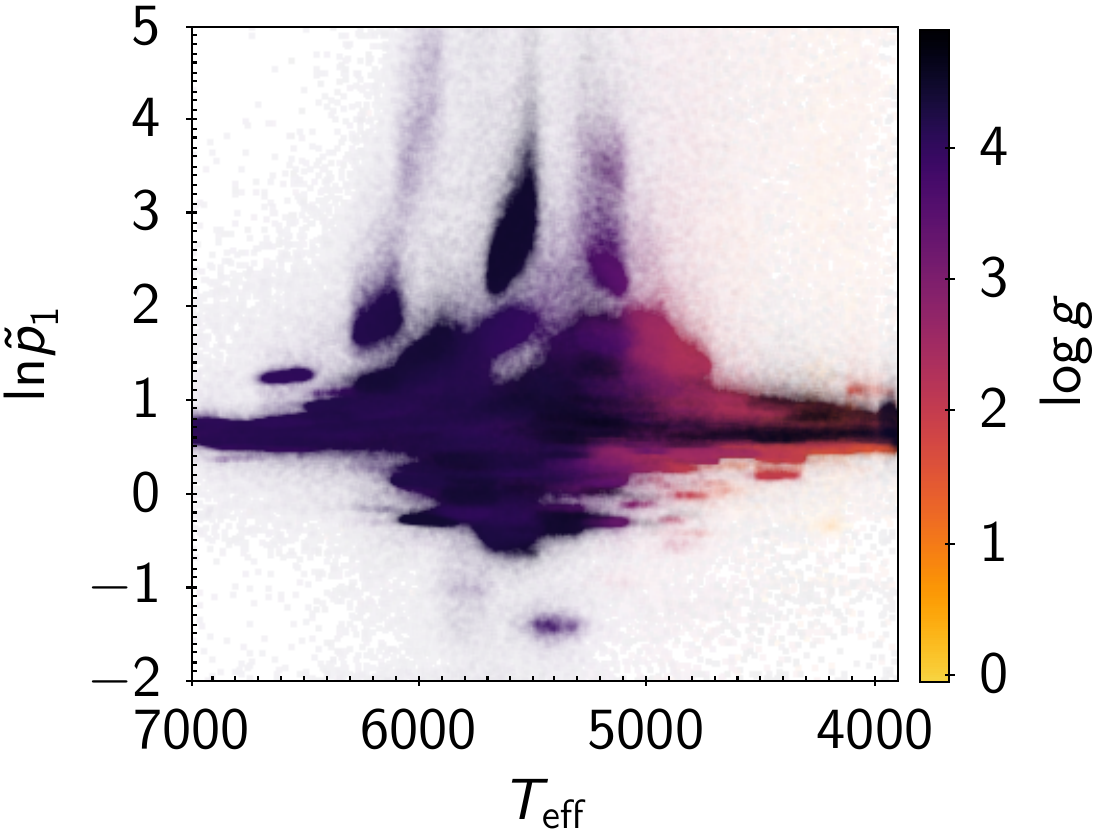}
    \includegraphics[width=0.45\linewidth]{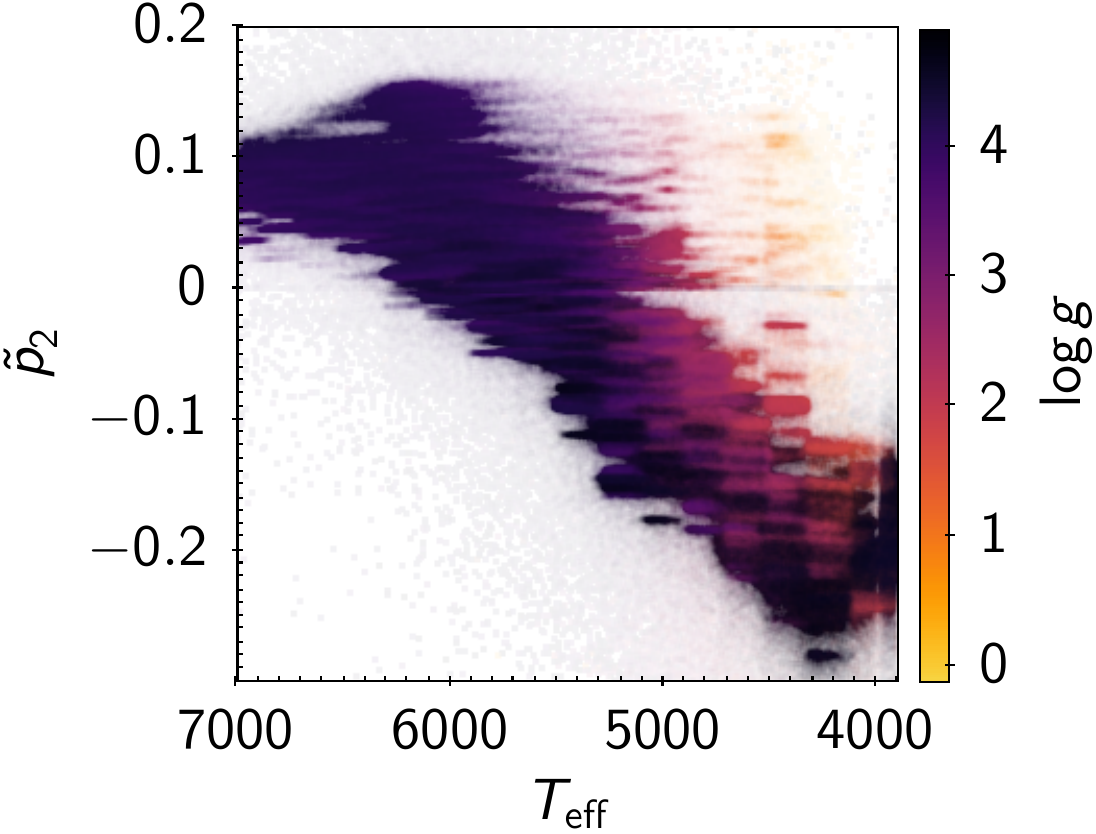}
    \caption{$T_\mathrm{eff}$ versus $p_1$ and $p_2$.}
    \label{teffp2}
\end{figure*}

\begin{figure*}
    \centering
    \includegraphics[width=0.9\linewidth]{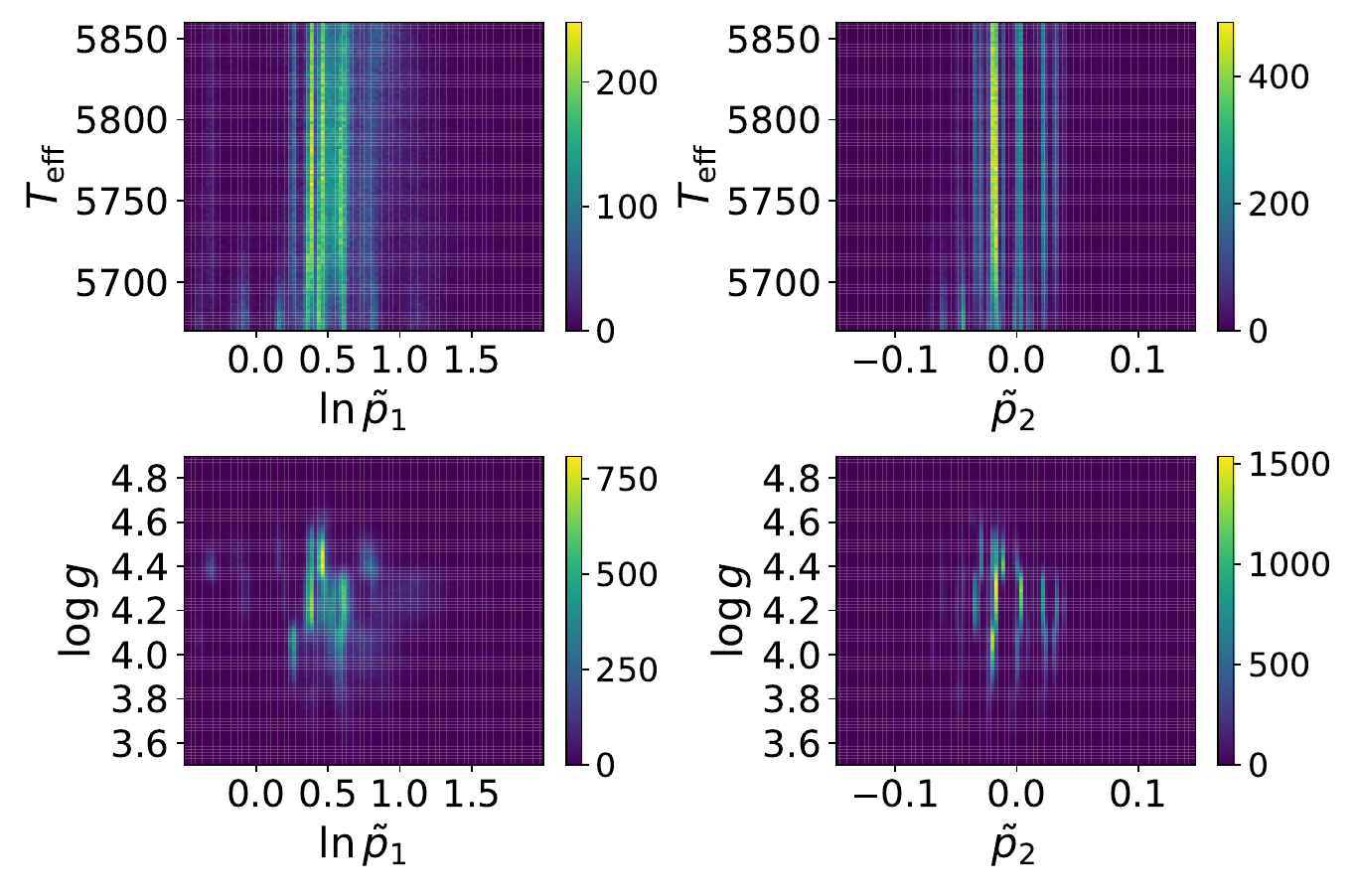}
    \caption{The relationship between $\tilde{p}_1$ and $\tilde{p}_2$ with respect to $T_\mathrm{eff}$ and $\log g$. $\tilde{p}_1$ and $\tilde{p}_2$ are from an example ($5670\mathrm{K} < {\it T}_\mathrm{eff} < 5860\mathrm{K}$) of the decoupled fitting parameters.}
    \label{decouple}
\end{figure*}

It is found that as the frequency of the spectrum increases, the spectral absorption depth becomes more sensitive to changes in the abundances of the neutron capture elements. Therefore, Eq.~\ref{1} is selected to fit the difference spectrum between the actual spectrum and the baseline spectrum.
\begin{equation}\label{eq}
    \Delta F=p_2 \cdot p_1^{-(x/1000-4)}
\end{equation}
Where the unit of x is \AA, and the wavelength range of spectral is 4000\AA to 8000\AA. The fitting parameters must contain information about neutron capture elements, but they must also include other information. As shown in Fig.~\ref{manyelements}, for solar-like stars, the fitting parameters are sensitive to Ca, Sc, Ti, Cr, Mn, Fe, Co, Ni, Sr, Zr, La, Ce, Pr, Nd, Sm, Gd, and Dy. The change in the other elements will not affect the fitting parameters too much. So, $p_1$ and $p_2$ can be regarded as the projections of atmospheric parameters, abundances of neutron capture elements, and several other elements in this specific two-dimensional parameter space. Among atmospheric parameters, $T_\mathrm{eff}$, $\log g$, and microturbulence have no direct relationship with the substructures in the Galactic disk, and should be decoupled from $p_1$ and $p_2$ before analysis. The specific decoupling process is as follows:

First, the median spectrum of one thousandth of the stars randomly selected from the LAMOST LRS Stellar Parameter Catalog of A, F, G and K Stars is calculated. One thousandth of the spectra are chosen randomly instead of the entire catalog to reduce computational costs. The median spectrum is chosen, instead of the average spectrum which has a lower computational cost, to exclude the impact of outlier spectra. The $p_1$ and $p_2$ distribution of the entire catalog is shown in Fig.~\ref{allteff}. It can be seen that $p_1$ and $p_2$ are coupled with $T_\mathrm{eff}$, which means that the same cluster is split into multiple clusters due to different $T_\mathrm{eff}$. Decoupling of the $T_\mathrm{eff}$ must be performed. Selecting a narrow temperature range can solve the problem, as shown in Fig.~\ref{teffp2}. Fig.~\ref{teffp2} is a deformation of Fig.~\ref{allteff}, which shows that $p_1$ and $T_\mathrm{eff}$ are not coupled, $p_2$ and $T_\mathrm{eff}$ are strongly coupled, however, in small intervals, $p_2$ and $T_\mathrm{eff}$ are almost decoupled. Besides, $\log g$ causes both $p_1$ and $p_2$ to stratify. As a result, we separate the LAMOST sample in to 16 $T_\mathrm{eff}$ ranges: 3900K, 4120K, 4340K, 4520K, 4710K, 4900K, 5090K, 5290K, 5460K, 5670K, 5860K, 6050K, 6240K, 6420K, 6680K, 6860K, 7000K, and recalculate the fitting parameters within each range. Only the main sequence star is retained, there are two reasons: On the one hand, remove the giants and sub-giants can decouple the fitting parameters and $\log g$. On the other hand, some mixed giants can transport the elements produced in their core onto the surface. Fig.~\ref{decouple} shows the distribution of the fitting parameters versus $T_\mathrm{eff}$ and surface gravity of the sample between 5670K and 5860K, demonstrating that decoupling has indeed been achieved. The decoupled fitting parameters are written as $\tilde{p}_1$ and $\tilde{p}_2$.

Among the known factors, microturbulence has not yet been decoupled from the fitting parameters so far. However, \cite{2004AN....325....3F} found that microturbulence has a strong correlation with $T_\mathrm{eff}$. Therefore, to a certain extent, $\tilde{p}_1$ and $\tilde{p}_2$ have been decoupled from the microturbulence. Even if the decoupling is not complete, the residual effect of microturbulence is unlikely to cause clustering in the distribution of $\tilde{p}_1$ and $\tilde{p}_2$.

To enhance our understanding of the interrelationships among data clusters, we performed a clustering analysis using the \texttt{HDBSCAN} algorithm from \texttt{sklearn.cluster} package. The clustering procedure was carried out in a two-dimensional parameter space, encompassing $\tilde{p}_1$, $\tilde{p}_2$. During this process, we employed the natural logarithm of pp1, and imposed constraints on $\ln \tilde{p}_1$ to reside between -4 and 70, and on $\tilde{p}_2$ to fall within the interval of -0.5 to 0.5, ensuring an absolute value greater than 10$^{-5}$ to preclude the inclusion of outliers. Stars with $\mathrm{SNR}_u < 10$ and the error of radial velocity larger than 10 (possible binaries) are also removed. Prior to the clustering analysis, the input parameters were normalized utilizing \texttt{StandardScaler} from the \texttt{sklearn.preprocessing} package, thereby facilitating a more robust evaluation of the data clusters.

\begin{figure*}
\centering
\includegraphics[width=0.42\linewidth]{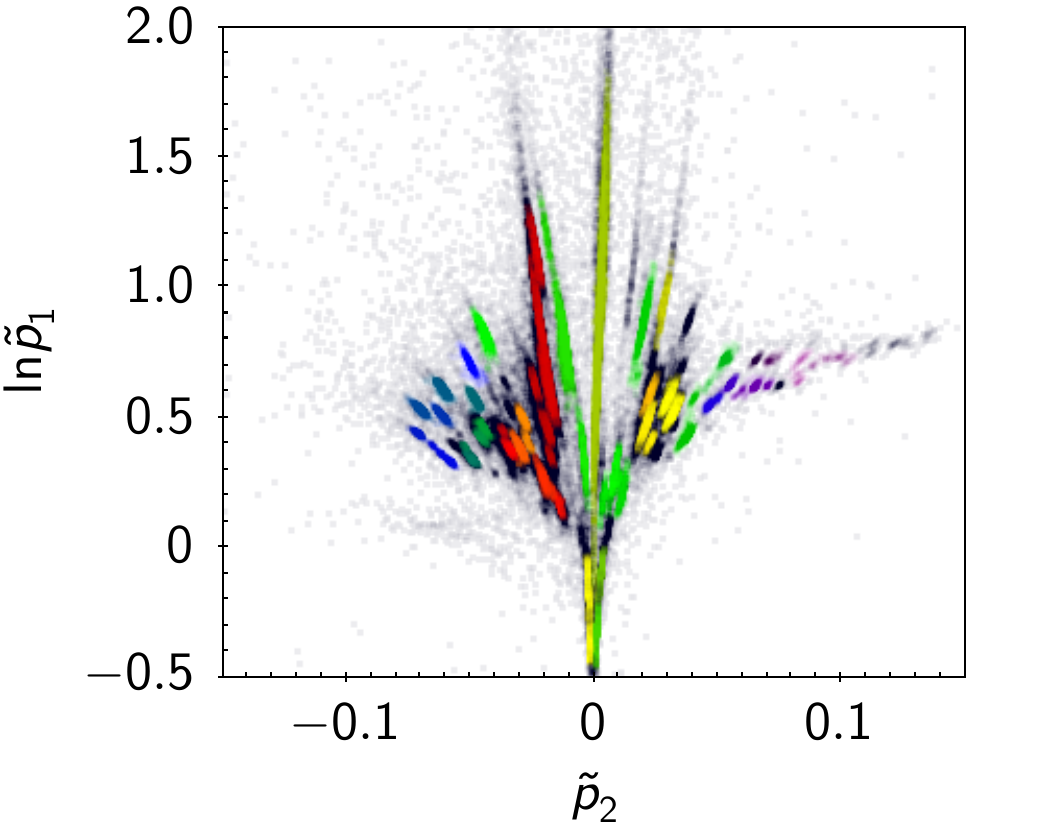}
\includegraphics[width=0.48\linewidth]{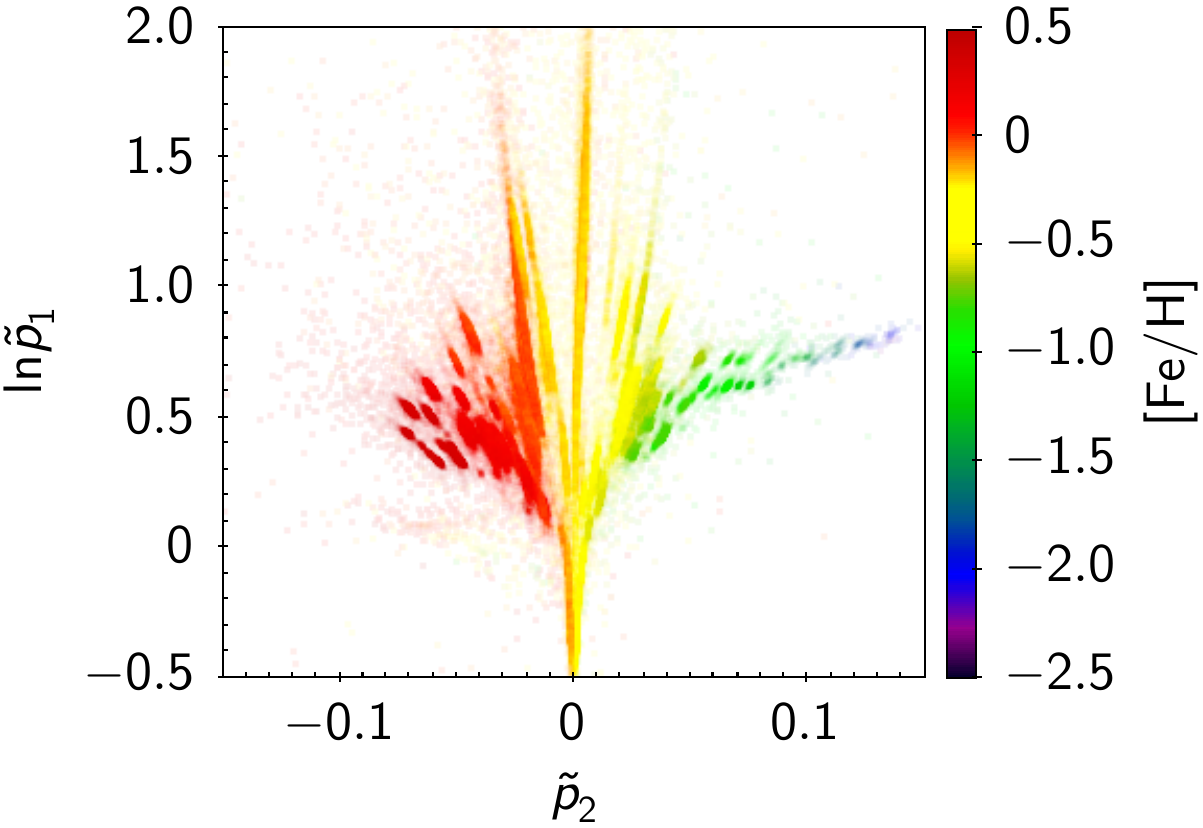}
\caption{The figure primarily depicts the central distribution of the fitting parameters for an example sample ($5670\text{K} < {\it T}_\mathrm{eff} < 5860\text{K}$) analyzed in this work. In the left panel, the color represents the HDBSCAN labels, gray means noise. In the right panel, the color represents [Fe/H].}\label{resultexample}
\end{figure*}

\begin{figure*}
\centering
\includegraphics[width=0.48\linewidth]{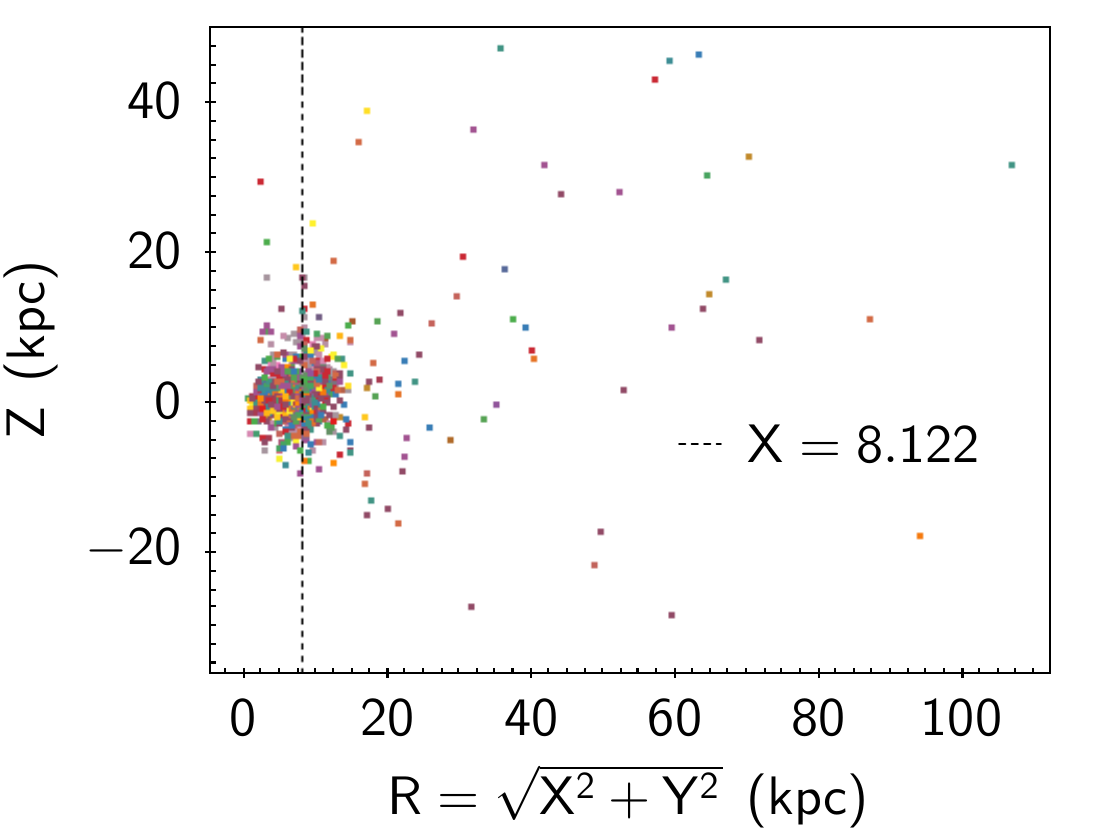}
\includegraphics[width=0.48\linewidth]{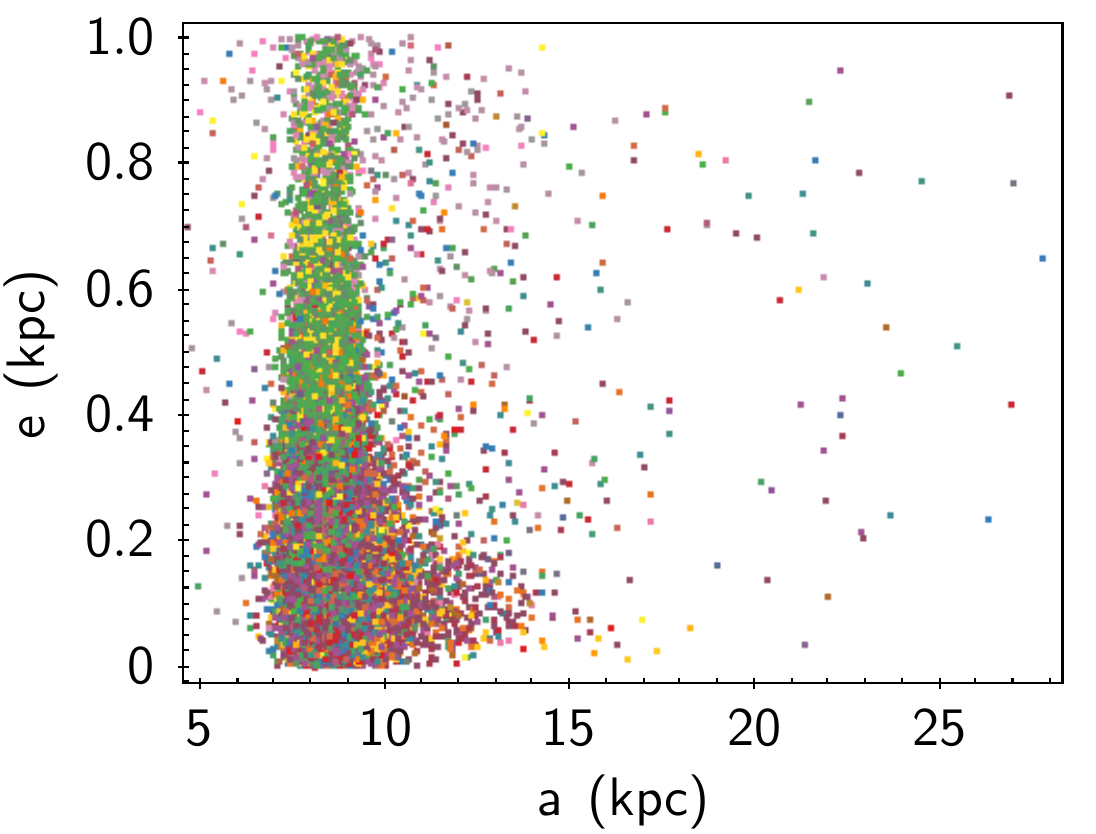}
\includegraphics[width=0.48\linewidth]{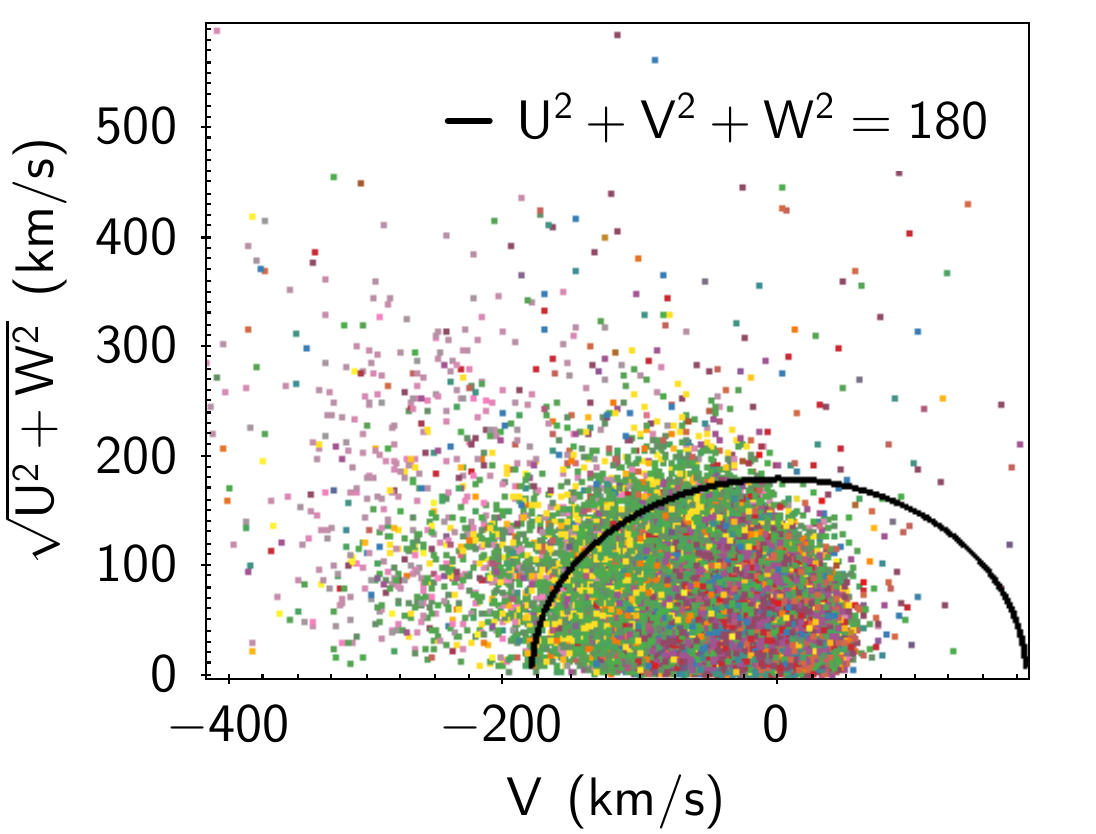}
\includegraphics[width=0.48\linewidth]{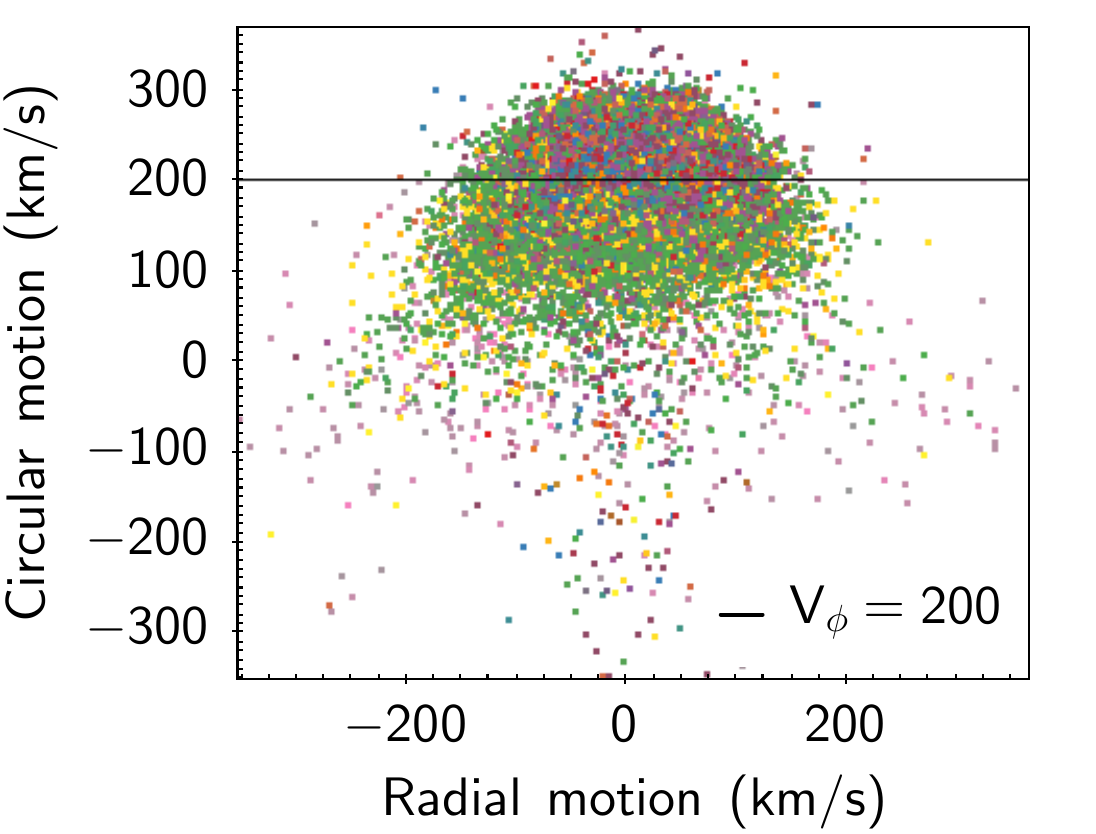}
\caption{The first panel depicts the Spatial distributions of the example sample ($5670\text{K} < {\it T}_\mathrm{eff} < 5860\text{K}$) analyzed in this work. The second panel depicts the distribution in the (a,e) space. The third and the forth panel depicts the distribution in the velocity space, but the coordinate systems are different. The color represents the cluster given by HDBSCAN in $\tilde{p}_1$ and $\tilde{p}_2$ space.}\label{kine}
\end{figure*}

Our investigation was particularly focused on two hyperparameters, namely \texttt{min\_cluster\_size} and \texttt{min\_samples}. For \texttt{HDBSCAN} method, the hyperparameters significantly influence the outcomes of clustering, evaluating the clustering results under different hyperparameters can be achieved visually. To improve the classification performance, in each $T_\mathrm{eff}$ range, the data are manually separated into two parts, low density and high-density areas. A grid search was executed to ascertain the optimal pairing of these hyperparameters, considering the values for \texttt{min\_cluster\_size} within the range of [400,300,200,100,50,40,30] for low density areas; [600,500,400,300,200,100] for high density areas. For \texttt{min\_samples}, both low and high density areas are set within [9, 7, 5, 3, 2, 1]. Upon applying the best hyperparameters in each $T_\mathrm{eff}$ interval, 16, 23, 18, 24, 28, 26, 48, 45, 50, 50, 61, 40, 58, 51, 38, 28 clusters are found. For colder stars, their selection effect is strong; For hotter stars, their sample size is small, both factors can lead to cluster loss. An example of clusters is shown in Fig.~\ref{resultexample}. Fig.~\ref{kine} depicts that the clusters do not have obvious kinematic features.


\begin{figure}
    \centering
    \includegraphics[width=0.9\linewidth]{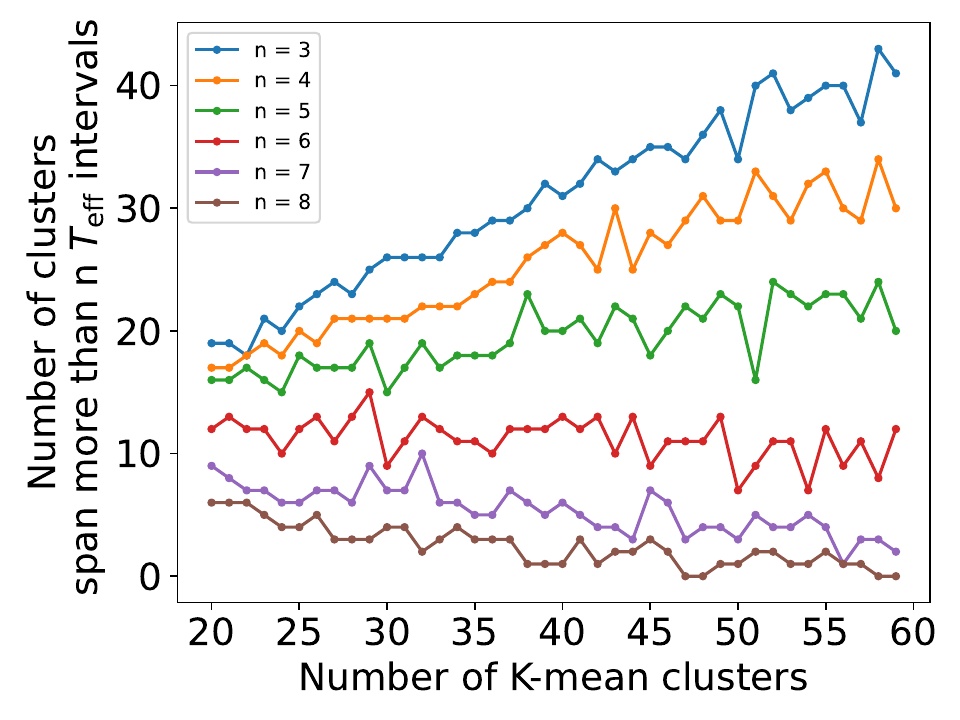}
    \caption{The results of the K-mean clustering of 538 \texttt{HDBSCAN} clusters in average [Fe/H], [Fe/H] scatter, average [$\alpha$/M], and [$\alpha$/M] scatter parameter space.}
    \label{showk}
\end{figure}

Among the 604 clusters we found, 538 of them have average [Fe/H], [Fe/H] scatter, average [$\alpha$/M], and [$\alpha$/M] scatter. K-means algorithm is performed in this four-dimensional parameter space. This time we focused on two hyperparameters, the first is the \texttt{number of clusters} required for K-means algorithm, which is set from 20 to 60; The second is the cluster which contains more than $n$ $T_\mathrm{eff}$ ranges, which is set from 3 to 8. The result is shown in Fig.~\ref{showk}. When $n=6$, the absolute slope of the number of clusters span more than n $T_\mathrm{eff}$ intervals and the number of K-mean clusters is the smallest, and there are about 10-12 clusters span more than 6 $T_\mathrm{eff}$ intervals.

\section{Discussions and Conclusions}\label{dis}

To gain a clearer understanding of the meanings of $\tilde{p}_1$, $\tilde{p}_2$, we conducted tests using more synthetic spectra. In these tests, all synthetic spectra had $T_\mathrm{eff}$=5500K, $\log g$=4.5, and [Fe/H] ranging from [-1.5,-1,-0.75,-0.5,-0.25,0,0.25,0.5]. The abundances of the elements were changed to [0.1,0.5,1,5,10] times the solar abundance, with the changed elements falling into three categories: 
\begin{description}  
    \item[Heavy]   
    Ga, Ge, As, Se, Br, Kr, Rb, Sr, Y, Zr, Nb, Mo, Tc, Ru, Rh, Pd, Ag, Cd, In, Sn, Sb, Te, I, Xe, Cs, Ba, La, Ce, Pr, Nd, Pm, Sm, Eu, Gd, Tb, Dy, Ho, Er, Tm, Yb, Lu, Hf, Ta, W, Re, Os, Ir, Pt, Au, Hg, Tl, Pb, Bi, Po, At, Rn, Fr, Ra, Ac, Th, Pa, U, Np, Pu, Am, Cm, Bk, Cf, Es  
  
    \item[S-only]   
    Mo, Tc, Ru, Rh, Pd, Ag, Cd, In, Sn, Sb, Te, I, Xe, Cs, Ba, La, Ce, Pr, Nd, Pm, Sm, Eu, Gd, Tb, Dy, Ho, Er, Tm, Yb, Lu, Hf, Ta, W, Re, Os, Ir, Pt, Au, Hg, Tl, Pb  
  
    \item[Alpha]   
    Mg, Si, Ca, Ti  
\end{description} 
Therefore, we obtained $8 \cdot 5 \cdot 3=120$ synthetic spectra. Subsequently, we added random noise to each synthetic spectrum, with the noise following a Gaussian distribution and the variance being [0.1,0.01,0.001] of the normalized flux at each wavelength, corresponding to SNRs of [10,100,1000]. For each parameter combination, we generated 1000 spectra with noise. To ensure that the median spectrum's [Fe/H] approximates 0, reflecting the actual abundance distribution of stars in our sample, we generated three times more samples with [Fe/H] = 0.25 and two times more samples with [Fe/H] = 0.5.

It is worth noting that these SNRs do not correspond to the SNR of the original spectra, as all spectra in this work were processed to a resolution of 10 nm. Considering that the noise at each point in the original spectra should not be independent, the SNR should increase significantly after reducing the resolution. This work limits the $\mathrm{SNR}_u >= 10$, and after reducing the resolution, the SNR of our sample should be significantly higher than 10.

\begin{figure*}
    \centering
    \includegraphics[width=0.3\linewidth]{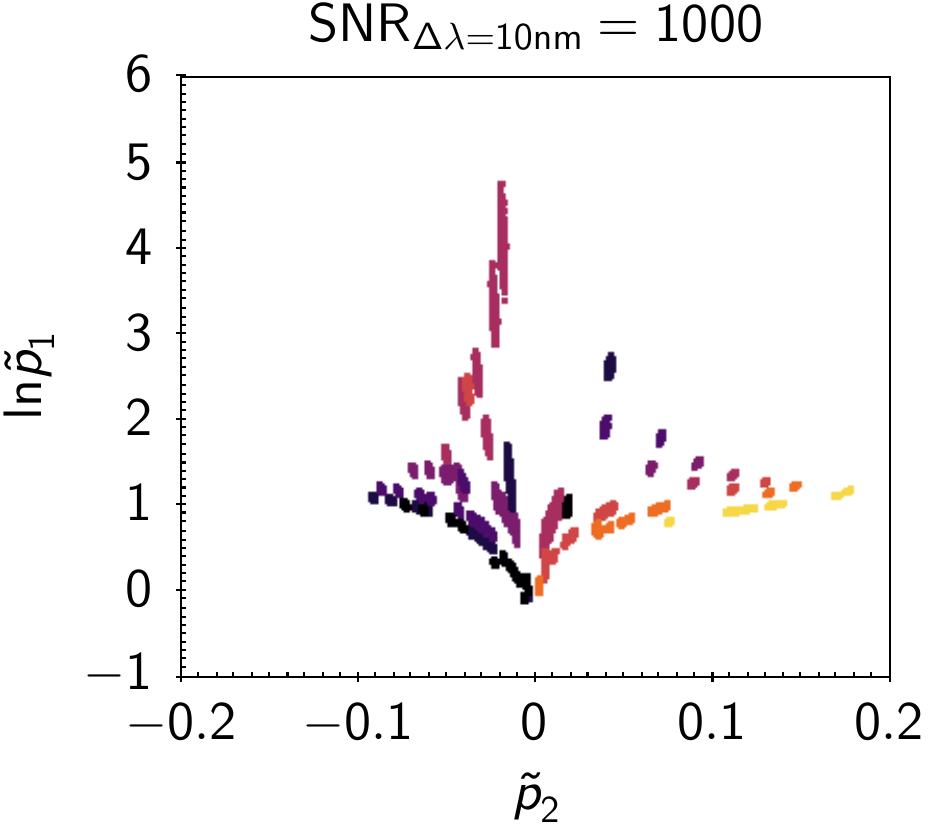}
    \includegraphics[width=0.3\linewidth]{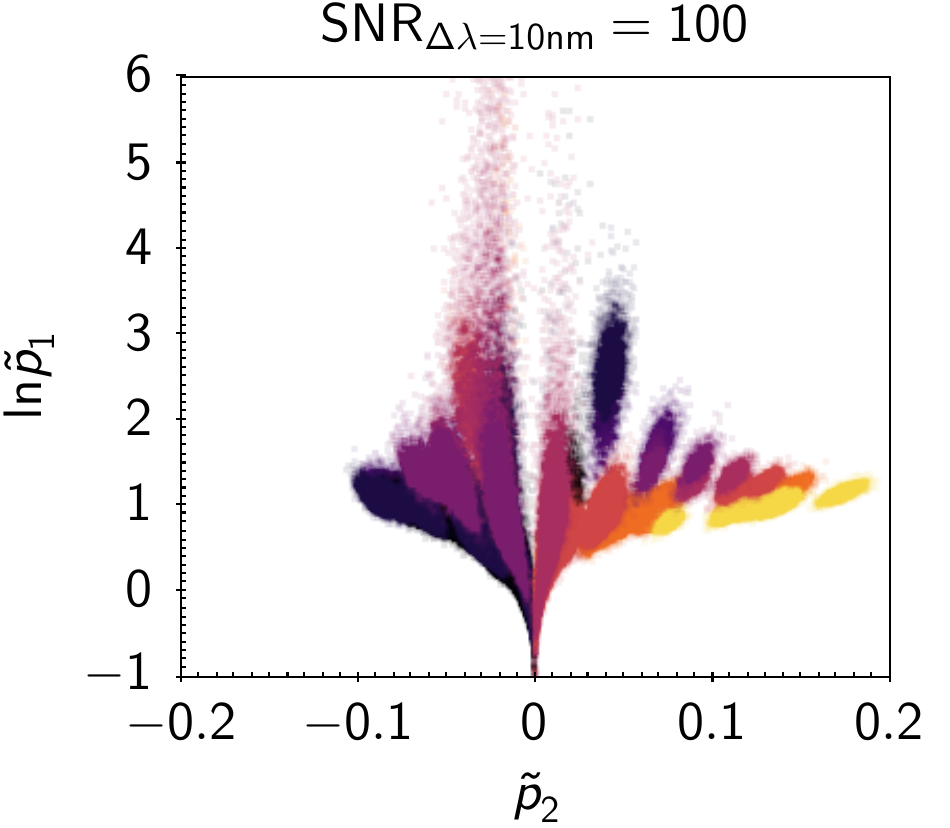}
    \includegraphics[width=0.355\linewidth]{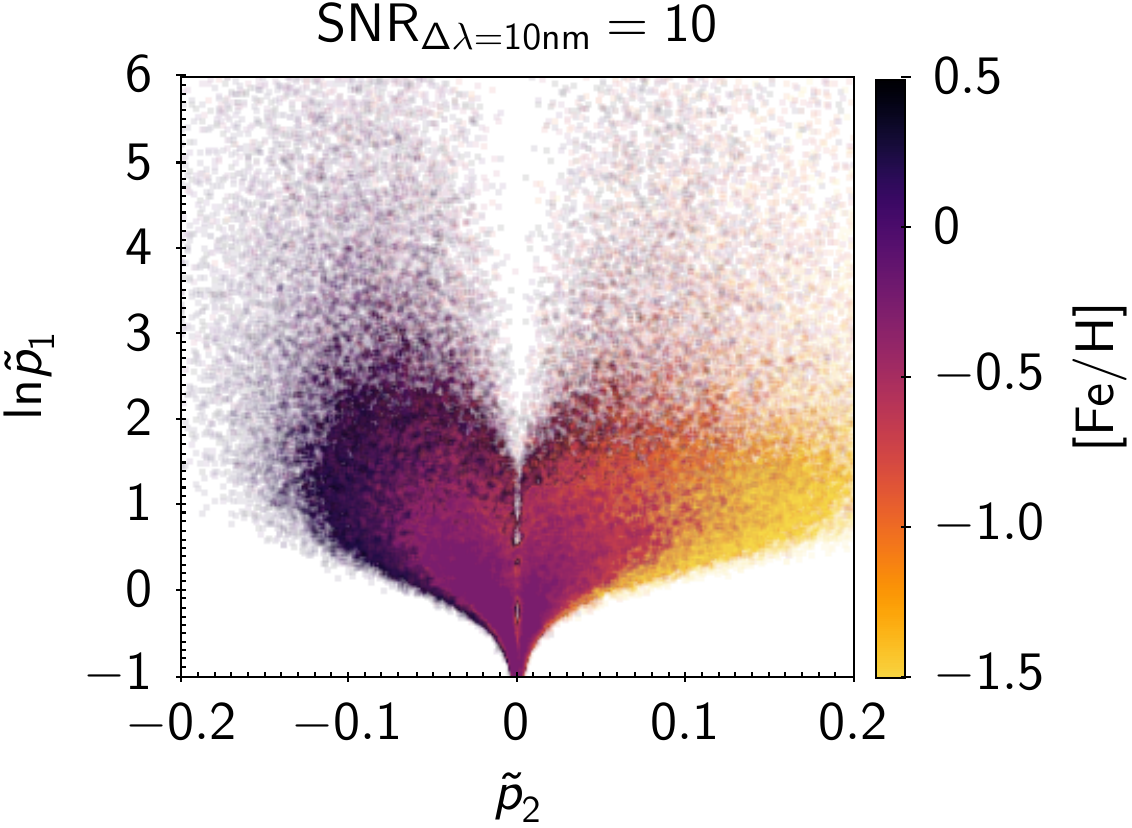}
    \caption{$\tilde{p}_1$ and $\tilde{p}_2$ distribution of synthetic spectra under different noise level. The colors represent [Fe/H].}
    \label{multiplesnrs}
\end{figure*}

As shown in Fig.~\ref{multiplesnrs}, under a resolution of 10 nm, no clusters can be distinguished when SNR = 10; obvious clusters can be identified when the SNR = 100, but they overlap significantly; and when SNR = 1000, not only can the clusters be clearly discerned, but also the three abundance variation patterns of ``heavy'', ``s-only'', and ``alpha'' can be distinguished within a certain range of parameters. However, this does not imply that $\tilde{p}_1$, $\tilde{p}_2$ can identify all three abundance patterns when the SNR is sufficiently high. There are two reasons for this: 1) The parameters of the synthetic spectra are discrete, while those of the actual spectra are continuous. 2) $\tilde{p}_1$, $\tilde{p}_2$ contain insufficient information since they are merely projections of a high-dimensional distribution onto this plane. Therefore, when using $\tilde{p}_1$ and $\tilde{p}_2$, only the lower limit of the number of clusters can be provided.

\begin{figure*}
    \centering
    \includegraphics[width=0.45\linewidth]{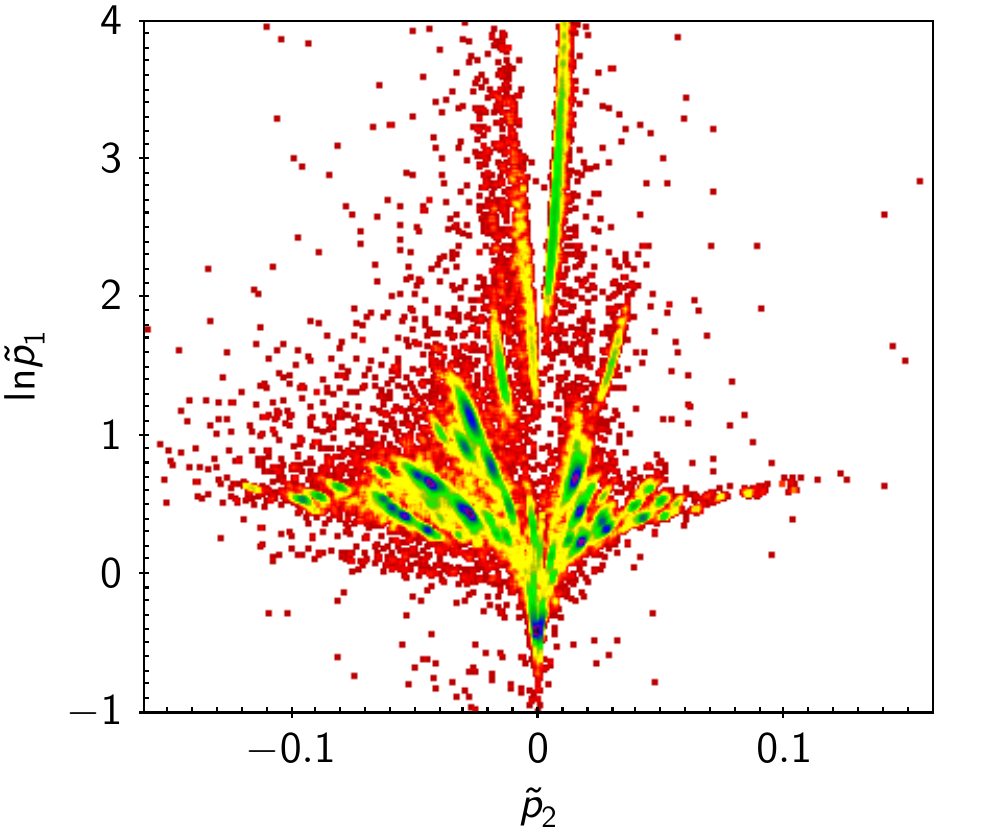}
    \includegraphics[width=0.45\linewidth]{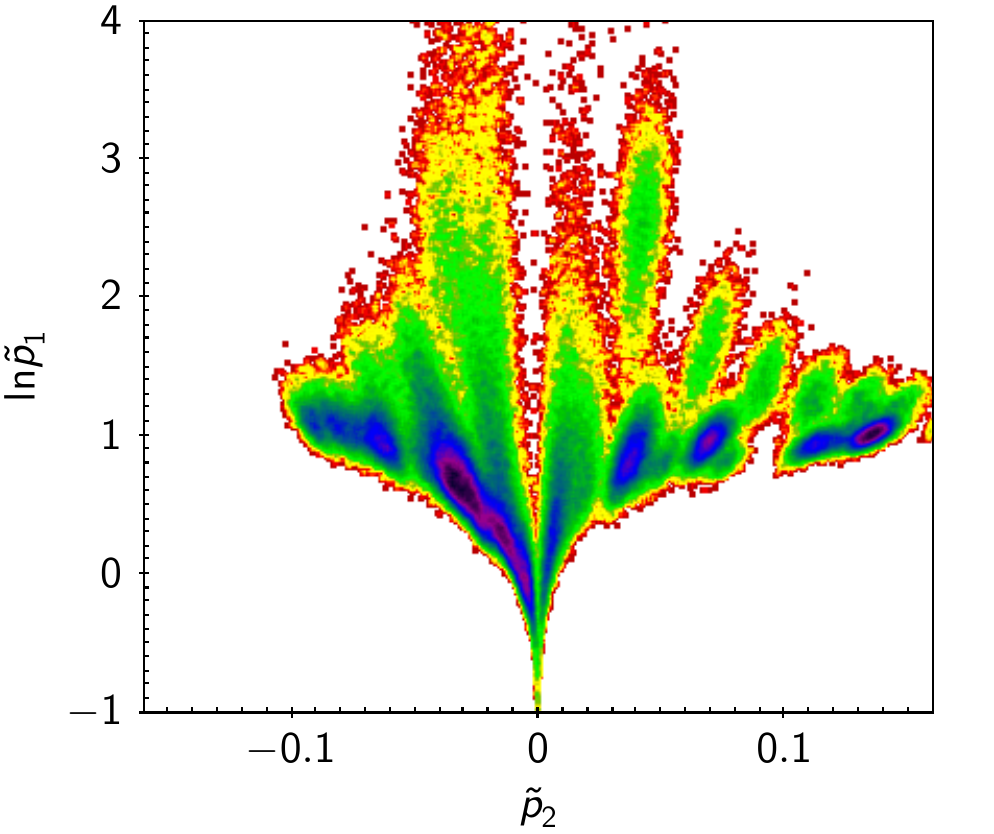}
    \caption{The left panel is the actual distribution from LAMOST DR9 ($5460\mathrm{K} < T_\mathrm{eff} < 5670\mathrm{K}$, main-sequence stars). The right panel is the synthetic distribution whose $T_\mathrm{eff} = 5500\mathrm{K}$ and $\text{SNR}_{\Delta \lambda=10 nm}=100$. The colors represent the density.}
    \label{syn_actual}
\end{figure*}

Fig.~\ref{syn_actual} compares the synthetic results of SNR = 100 with actual data ($5460\mathrm{K}< T_\mathrm{eff} < 5670\mathrm{K}$). It can be seen that compared with the synthetic distribution, the actual distribution has a smaller $\tilde{p}_1$. This is because the continuous spectrum of the synthetic data is a perfect black-body spectrum, which is impossible for actual data. LAMOST lacks absolute flux calibration and has a lower resolution, resulting in a poorer quality of the continuous spectrum compared to the synthetic spectrum, making the normalized spectrum flatter. This leads to deviations in $\tilde{p}_2$. However, the impact of this effect on $\tilde{p}_1$ is limited. As can be seen from Eq.~\ref{eq}, $\tilde{p}_2$ is much more sensitive to the data than $\tilde{p}_1$.

\begin{acknowledgments}

\end{acknowledgments}

%


\software{TOPCAT \citep{2005ASPC..347...29T},
            astropy \citep{2013A&A...558A..33A,2018AJ....156..123A,2022ApJ...935..167A}
          }






\bibliography{sample631}{}
\bibliographystyle{aasjournal}



\end{document}